# Plasmon-enhanced Stimulated Raman Scattering Microscopy with Single-molecule Detection Sensitivity

Cheng Zong[1,2], Ranjith Premasiri[3,4], Haonan Lin[1], Yimin Huang[3], Chi Zhang[1], Chen Yang[3], Bin Ren[2], Lawrence D. Ziegler[3,4], and Ji-Xin Cheng[1,3,4]*

1 Department of Electrical and Computer Engineering, Department of Biomedical Engineering, Boston University, Boston, Massachusetts 02215, USA.

2 State Key Laboratory of Physical Chemistry of Solid Surfaces, MOE Key Laboratory of Spectrochemical Analysis and Instrumentation, Collaborative Innovation Center of Chemistry for Energy Materials, College of Chemistry and Chemical Engineering, Xiamen University, Xiamen 361005, China.

3 Department of Chemistry, Boston University, Boston, Massachusetts 02215, USA.

4 Photonics Center, Boston University, Boston, Massachusetts, 02215, USA.

**Abstract**

Stimulated Raman scattering (SRS) microscopy allows for high-speed label-free chemical imaging of biomedical systems. The imaging sensitivity of SRS microscopy is limited to ~10 mM for endogenous biomolecules. Electronic pre-resonant SRS allows detection of sub-micromolar chromophores. However, label-free SRS detection of single biomolecules having extremely small Raman cross-sections (~$10^{-30}$ $cm^2$ $sr^{-1}$) remains unreachable. Here, we demonstrate plasmon-enhanced stimulated Raman scattering (PESRS) microscopy with single-molecule detection sensitivity. Incorporating pico-Joule laser excitation, background subtraction, and a denoising algorithm, we obtained robust single-pixel SRS spectra exhibiting the statistics of single-molecule events. Single-molecule detection was verified by using two isotopologues of adenine. We further demonstrated the capability of applying PESRS for biological applications and utilized PESRS to map adenine released from bacteria due to starvation stress. PESRS microscopy holds the promise for ultrasensitive detection of molecular events in chemical and biomedical systems.

**Introduction**

Raman spectroscopy is a versatile analytical tool providing information about the native fingerprint vibrational states of a sample determined by a molecule's structure and its environment. Non-electronically resonant spontaneous vibrational Raman scattering cross-sections are typically $10^{-30}$ $cm^2$ $sr^{-1}$ and intrinsically small cross-sections on this order result in detection limits only as low as milli-molar (mM) levels. By placing a molecule close to a plasmonic nanostructure, plasmon-enhanced Raman spectroscopy pushes the detection sensitivity to the single-molecule level,[1-9] yet the speed in spectral acquisition is still not sufficient for ultrasensitive chemical mapping of molecular events in a dynamic and complex system.[10]

Owing to the development of advanced lasers and electro-optic instruments, nonlinear Raman microscopy has been shown to provide label-free chemical imaging, based on either coherent anti-Stokes Raman scattering (CARS) or stimulated Raman scattering (SRS), for a broad range of biomedical applications.[11] Early developments of CARS and SRS microscopy relied on picosecond pulses for detection of a single Raman peak.[12, 13] Intra-pulse broadband CARS, developed by Cicerone and coworkers, allowed recording of a whole Raman spectrum within 3.5 ms.[14] Multiplex SRS microscopy developed by Cheng and coworkers[15], is able to acquire a Raman spectrum covering a 200 wavenumber spectral window within 5 µs, which allowed high-throughput chemical analysis in a flow cytometry setting.[16] Yet, the imaging sensitivity of SRS microscopy is limited to ~10 mM for chemical bonds such as the C-H vibrations in cell membranes.[17, 18]. Min and coworkers recently reported electronic pre-resonance SRS achieving sub-µM-sensitivity detection for chromophores having a Raman cross-section over $10^3$ or $10^4$ times larger than endogenous biomolecules.[19, 20]

To push coherent Raman detection sensitivity further, plasmon-enhanced CARS has been reported [21-26] and single-molecule sensitivity has been proved.[23, 24] While, the CARS signal carries a non-resonance background which complicates quantification and distorts the CARS spectrum. Additionally, the CARS signal displays a nonlinear dependence on the concentration of analytes.[27] The SRS signal, on the other hand, exhibits identical spectral profile as spontaneous Raman and a linear dependence on the concentration of analytes.[13] The Van Duyne group reported reproducible surface-enhanced femtosecond SRS spectra from molecules embedded in a gold nano-dumbbell sol.[28, 29] Yet, plasmon-enhanced SRS at single-molecule detection sensitivity has not been reported.

Major hurtles of achieving single-molecule SRS detection include the damage of plasmonic substrates by the ultrafast pulses[30] and a large pump-probe background, arising from plasmon-induced photothermal and/or stimulated emission process.

Here, we report plasmon-enhanced SRS (PESRS) microscopy (**Fig 1a**, instrument in **SI1**) and its application to ultrasensitive imaging of biomolecules released from a cell. We reached single-molecule detection sensitivity by incorporating several innovations. First, we used chirped laser pulses at 80 MHz repetition rate for spectral-focusing hyperspectral SRS imaging. The pulse energy on the sample was on the level of pJ. Such low pulse energy together with chirping to picosecond duration effectively avoided sample photodamage, while the high repetition rate allowed fast chemical mapping of molecules adsorbed on gold nanostructured surfaces. Second, we employed a penalized least squares (PLS) approach and successfully extracted the sharp Raman peaks from a spectrally broad non-Raman background largely contributed by the photothermal effect.[31] Third, harnessing a block-matching and 4D filtering (BM4D) algorithm to denoise a hyperspectral stack, we were able to generate high-quality single-pixel SRS spectra for statistical analysis of single-molecule events. By a bianalyte method,[32-35] we used two isotopologues of adenine that offer unique vibrational signatures and verified PESRS detection of single molecules with Raman cross-section as low as $10^{-30}$ cm$^2$ per molecule. Furthermore, we demonstrated PESRS imaging of adenine resulting from nucleotide degradation as a stress response of *S. aureus* cells to starvation.

**Results**

**Plasmon-enhanced stimulated Raman scattering (PESRS) spectroscopy.** Adenine adsorbed on Au NPs aggregation substrates (see Methods) was selected as a proof-of-principle system for the demonstration of PESRS. Adenine is one of the four constituent bases of nucleic acids. The Raman band at 723 cm$^{-1}$ of adenine powder, which has a cross-section of 2.9×10$^{-30}$ cm$^2$,[36] has been studied for a single-molecule detection by surface-enhanced Raman spectroscopy (SERS)[35, 36] and surface-enhanced CARS.[23] As shown in **Fig. 1a**, a pump laser centered at 969 nm and a Stokes laser centered at 1040 nm were employed to induce a PESRS spectrum covering a window ranging from 550 to 850 cm$^{-1}$. 10 µL of a 5 mM aqueous adenine solution was added to 2~4 µL of a concentrated Au colloid suspension which induced the aggregation of Au NPs. A representative extinction spectrum of an adenine-induced Au NPs aggregation substrate is shown in **Fig. 1b**. The plasmonic band of the aggregated Au NPs is broad and peaked at 1040 nm, which allows PESRS for the pump and Stokes laser wavelength used here. The resulting PESRS spectrum (**Fig. 1c**, black) from the adenine-adsorbed Au NPs aggregates consists of a narrower feature at 733 cm$^{-1}$ (highlighted by green) on top of a strong and broad non-resonant background. This sharp feature is close to the prominent adenine ring-breathing mode frequency observed in the normal SRS spectrum of adenine powder (**Fig. 1c**, blue) and identical to the corresponding 733 cm$^{-1}$ peak observed in the SERS results on Au substrates (**SI2**).[37] The blank result (**Fig. 1c**, red) was independently measured from the Au NPs substrate without adenine adsorption. The background could arise from three different non-Raman processes: a photothermal effect, cross-phase modulation, and transient absorption,[31] all due to laser interactions with the Au nanostructures. The spectral shift between the substrate with/without adenine may related to the different extent of aggregation with/without adenine. These backgrounds are spectrally overlapped with the SRS signal, but are largely independent of the Raman shift.[31] In contrast, the SRS signal originates from a vibrational resonance that has a sharp spectral feature. A PLS approach was used to fit the broad spectral background. The resulting fitting backgrounds of PESRS are shown in **Fig. 1c** as the same color dash lines for the corresponding observed PESRS spectra. **Fig. 1d** shows the vibrationally resonant component of the PESRS spectra resulting from subtraction of the fitting backgrounds from the observed PESRS signals. The PESRS spectrum of adsorbed adenine shows a dominated peak at 733 cm$^{-1}$. Only a noisy baseline is evident after background subtraction from the pure substrate spectrum. Compared with the SRS spectrum of adenine crystal (blue line in **Fig. 1d**), a 10 cm$^{-1}$ blue shift of the peak is observed in the PESRS spectrum. This blue shifted frequency (733 cm$^{-1}$) is consistent with the strongest vibrational feature observed in SERS spectra of adenine (**SI2**).[38] These results collectively indicate that the observed vibrational PESRS signal component originates from the surface adsorbed adenine. **Fig. 1d inset** presents the standard SRS spectrum of adenine powder at the same laser power condition as used for the detection of PESRS. The standard SRS setup could not generate any Raman signal from a pure adenine powder, while PESRS could detect a thin layer of adenine adsorbed on Au nanostructures. This result indicates that the large electromagnetic field boosted by the plasmon significantly amplified the stimulated Raman process.

To verify that the SRS signal is due to the adenine vibrational resonance, we varied the pump wavelength while keeping the Stokes wavelength fixed. The pump laser centered at 972 nm as well as the previous 969 nm wavelength encompass the adenine Raman resonance for a 1040 nm Stokes pulse, and both generated SRS spectra showing a pronounced peak at the expected wavenumber (**Fig. 1e**, black and red). In contrast, the 942 nm is off-resonance for the 733 cm$^{-1}$ band. Accordingly, the measured spectrum does not exhibit such a peak as shown in **Fig. 1e** (blue). After subtraction of the background in **Fig. 1e** (corresponding fitted backgrounds were shown as same color dash lines), the PESRS spectra of adenine excited by both Raman resonance wavelengths show a Raman peak at 733 cm$^{-1}$ (**black and red, Fig. 1f**), whereas the off-resonance spectrum only shows a noisy featureless baseline (**blue, Fig. 1f**). Moreover, as shown in **SI3**, the intensity of the 733 cm$^{-1}$ peak linearly depends on the pump power and the Stokes power before it reaches saturation. The reproducible spectra recorded at the same location demonstrate that the laser power in our experiment did not damage the substrate or induce molecular photodegradation. These results collectively confirm the SRS origin of the vibrationally resonant component of the observed spectrum and the plasmonic enhancement of this signal. To ensure that our method is not specific to adenine, we tested other molecules. **SI4** shows the PESRS spectra of Rhodamine 800 (85 μM in solution) and 4-mercaptopyridine (5.7 mM in solution) adsorbed on the Au NPs aggregated substrate.

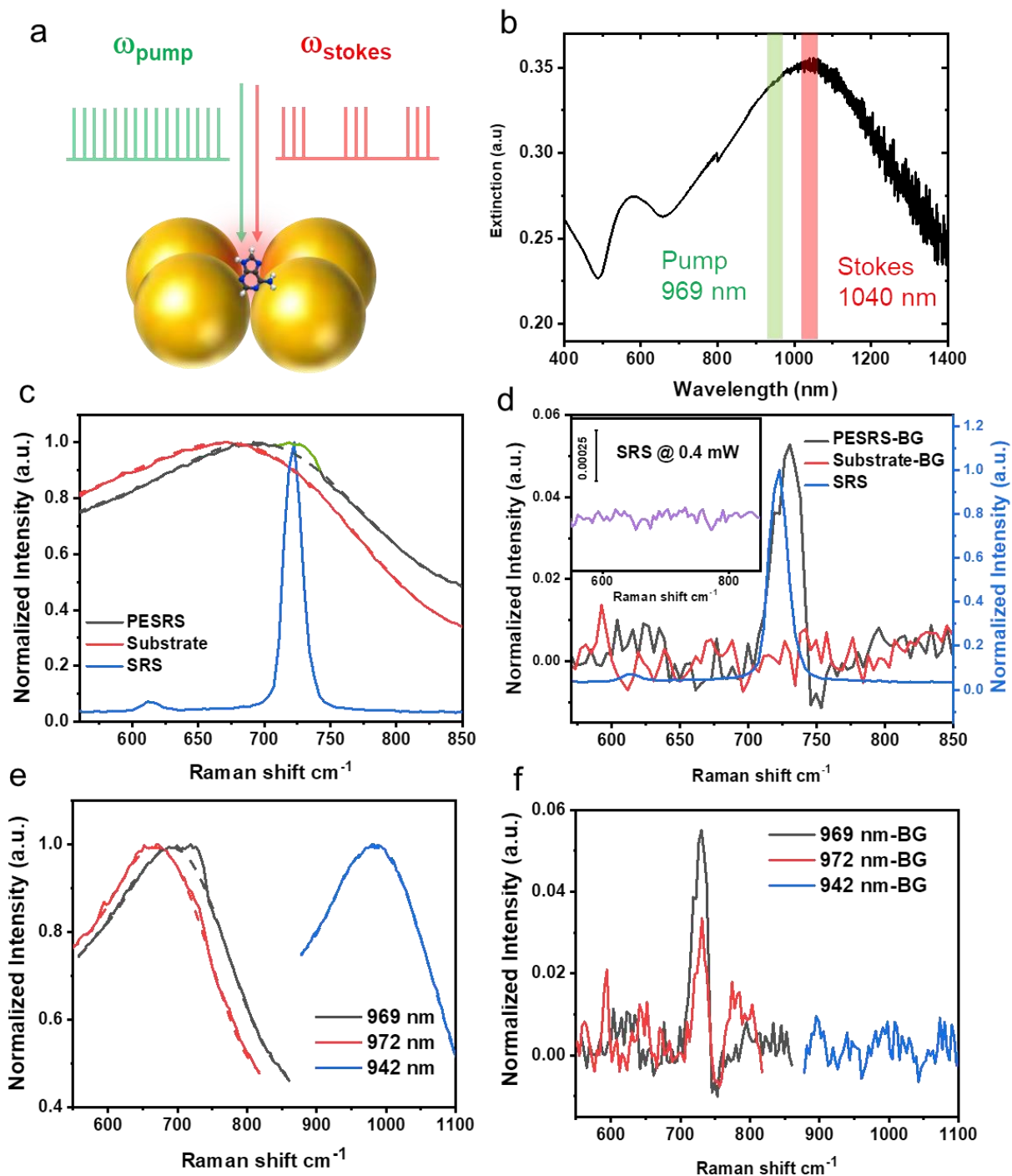

**Figure 1. PESRS spectroscopy**. (**a**) Schematic. (**b**) A representative extinction spectrum of adenine-induced Au NPs aggregation substrate. (**c**) The PESRS spectrum (solid) with a green highlighted portion and fitted background (dash) obtained from the substrate with adenine adsorption. The blank spectrum (solid) and fitted background (dash) obtained from the substrate without adenine. The total power of pump and Stokes was 0.4 mW. The SRS spectrum of adenine powder (blue) was obtained with a pump power at 10 mW and a Stoke power at 50 mW. (**d**) The background-subtracted PESRS spectrum of adsorbed adenine

versus the SRS spectrum of adenine powder (same as blue line in c) and the spectrum of blank substrate. Inset: The SRS spectrum of adenine powder obtained as the same laser power condition as the PESRS. (**e**) PESRS spectra (solid) and fitted background (dash) of adenine at Raman resonance (969 nm and 972 nm) and off-resonance (942 nm). (**f**) Background-subtracted PESRS spectra of adenine at Raman resonance and off-resonance. BG: background.

**PESRS at single-pixel level.** To demonstrate the imaging capability of PESRS, we scanned the adenine containing aggregated Au NPs substrate with a pixel dwell time of 10 μs. It took c.a. 1 min to obtain a hyperspectral cube (200 × 200 pixel, and 80 Raman channels) consisting of 40000 spectra. In **Fig. 2a**, the averaged total 80 spectral channels of an original PESRS hyperspectral data cube are plotted to show the spatial distribution of aggregated NPs. **Fig. 2b** shows two single-pixel spectra from regions with and without NP aggregates, indicated as spot 1 and spot 2, respectively. The single-pixel spectrum from spot 1 shows a broad background and a weak Raman peak around 733 cm$^{-1}$. After pixel by pixel subtraction of the fitting background, the area of the resulting vibrational band at 733 cm$^{-1}$ at each pixel is shown in **Fig. 2c** revealing a clear spatial contrast between regions of adsorbed adenine and blank areas. The single-pixel background-removed spectra from spot 1 and 2 are displayed in **Fig. 2d**. It remains challenging to obtain high-quality single-pixel spectra due to the noisy non-Raman background. To address this challenge, we employed a BM4D algorithm which was widely used for 3D data denoising.[39, 40] The reconstructed peak area image and the single-pixel spectra after BM4D denoising and background removal are shown in **Figs. 2e** and **2f**, respectively. By employing BM4D, we achieved a signal-to-noise ratio of 33 for the single-pixel spectra at spot 1, ~4 times better than that without denoising. Good Raman reproducibility in terms of peak frequency in different locations of the imaging area was found (see **SI5**) even given an expected inhomogeneous intensity distribution due to the Au NPs randomly aggregated hot spots.

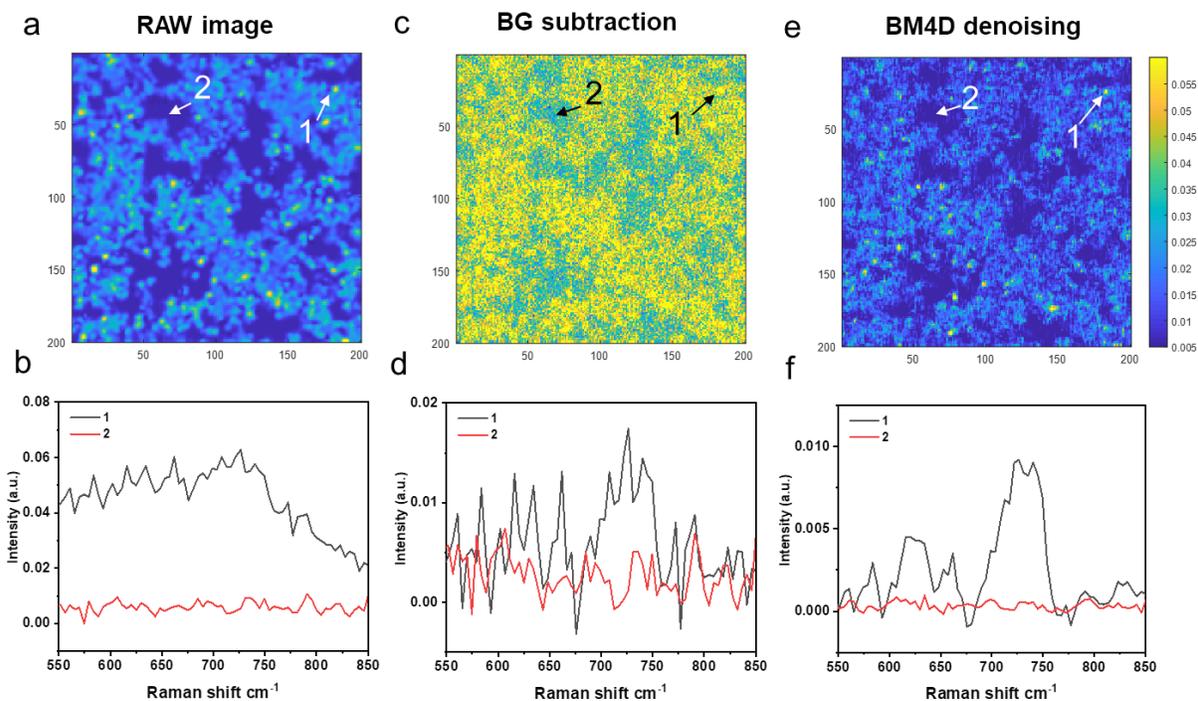

**Figure 2. Single-pixel PESRS.** (**a**) The raw PESRS image of aggregated Au NPs substrate with adsorbed adenine. The color of each pixel tracks the average total spectral channels intensity from each PESRS spectrum. (**b**) The raw single pixel spectra obtained from spot 1 and 2, which are indicated in (a). (**c**) The PESRS image of adsorbed adenine on aggregated Au NPs substrate. The color of each pixel tracks the peak area at 733 cm$^{-1}$ after background subtraction. (**d**) The background-removed single-pixel spectra obtained from spot 1 and 2 which are indicated in (c). (**e**) The BM4D-denoised PESRS image of adsorbed adenine on aggregated Au NPs substrate. The color of each pixel tracks the peak area at 733 cm$^{-1}$ after BM4D denoising and background subtraction. (**f**) The BM4D-denoised single-pixel spectra obtained from spot 1 and 2 which are indicated in (e). Image area: 30 μm × 30 μm.

PESRS was also demonstrated for epi-detection of molecules on a non-transparent plasmonic substrate, since such substrates are often used in plasmon-enhanced spectroscopy applications. The experimental setup is shown in **Fig. 3a**. We used a sol-gel-derived SiO$_2$ substrate covered by immobilized aggregates of monodispersed-sized Au NPs (AuNPs-SiO$_2$ substrate).[41] 10 μL of a 100 μM adenine solution were dropped on this plasmonic substrate and dried the sample in air. The spectrally integrated image (**Fig. 3b**) reveals the distribution of NP clusters on the SiO$_2$ chip. After BM4D denoising and background subtraction, the distribution of hot spots is evident in **Fig.**

**3c**. Single-pixel spectra extracted from spots 1 and 2 are indicated in **Fig. 3d**. After denoising, a signal-to-noise ratio of 48 is achieved for these single-pixel spectra. These data collectively show the high sensitivity of epi-detected PESRS.

To estimate the relative enhancement factor of a local hot spot, we assumed a monolayer surface coverage of adenine and a monolayer NP cluster under the laser focus. Based on the measured local PESRS intensity (spot 1) and the average SRS intensity of 5 mM adenine solution, the power- and concentration-averaged local enhancement factor of PESRS relative to normal SRS is estimated to be ~ $7\times10^7$ (see details in **SI6**). Consistent with this result, enhancement factors of $10^4$ ~$10^6$ and $10^5$ ~$10^8$ were reported for surface-enhanced femtosecond SRS[28] and surface-enhanced CARS spectroscopy,[22, 23, 25] respectively.

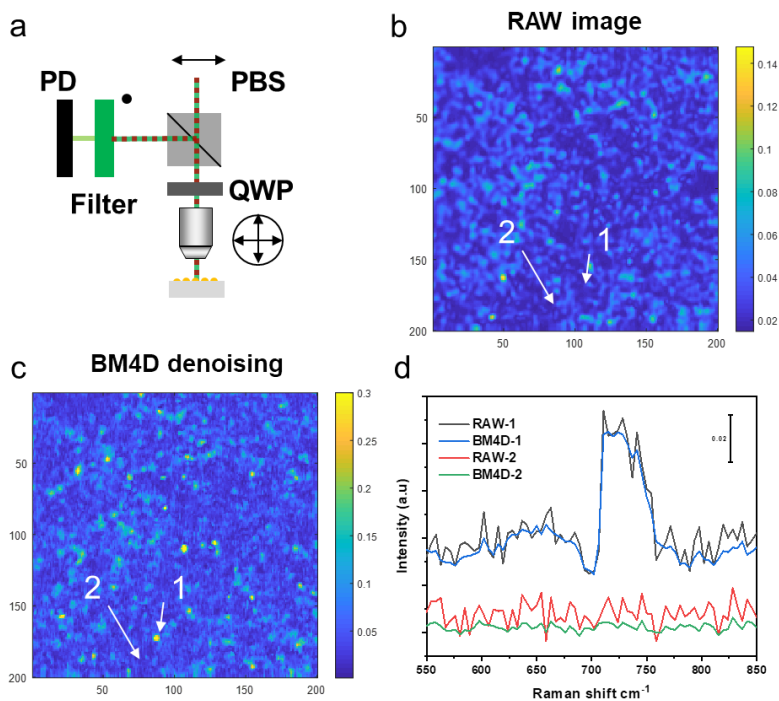

**Figure 3. Epi-detected PESRS.** (**a**) The schematic. A polarizing beam splitter (PBS) and a quarter wave plate (QWP) changes the polarization of incoming and backscattered lasers by 90°. In this way, the stimulated Raman loss signal passes the filter and is detected by a photodiode (PD). (**b**) The raw PESRS image of adenine adsorbed on Au NPs-SiO$_2$ substrate. The color of each pixel tracks the average intensity

from each PESRS spectrum. (**c**) The denoised PESRS image of adenine adsorbed on Au NPs-SiO$_2$ substrate. The color of each pixel tracks the peak area at 733 cm$^{-1}$ from each denoised and background-corrected PESRS spectrum. The whole image area is 30 μm × 30 μm. (**d**) The single-pixel spectra of adenine on the Au NPs-covered SiO$_2$ substrate obtained from spot 1 and 2 which indicate in (**c**).

**Single-molecule sensitivity in PESRS**. To quantitate the detection sensitivity of PESRS, we used the bianalyte approach developed by the Le Ru[32, 33] and Van Duyne[34] groups. The bianalyte approach relies on the observation of two different analytes adsorbed on a nanostructured surface. At the single-molecule level, components of the observed spectra can be attributed exclusively to one or the other of the two analytes by virtue of its distinguishable Raman spectrum. The most straightforward bianalyte approach uses a pair of isotopologues that are chemically identical but spectrally distinct. Here, we used a pair of isotopic molecules of adenine ($^{14}$NA) and adenine-1, 3-$^{15}$N$_2$ ($^{15}$NA). PESRS spectra of ensemble averaged pure $^{14}$NA and pure $^{15}$NA (both 1 mM in solution) show clearly distinguishable Raman bands centered at 733 cm$^{-1}$ and 726 cm$^{-1}$, respectively (**Fig. 4a**).[35, 36] The PESRS spectrum of an equimolar solution of $^{14}$NA and $^{15}$NA displays a single Raman band at 730 cm$^{-1}$. These frequency features allow for unequivocal identification of individual molecules and their mixture in PESRS spectra.

To evaluate the sensitivity of PESRS, we prepared a mixture solution of $^{14}$NA and $^{15}$NA at 500 nM concentration each with Au NPs and dried the colloid on a cover glass under vacuum. A hyperspectral PESRS image of this mixture sample, consisting of 40000 spectra, was acquired for statistical analysis. **Fig. 4b** shows typical single-pixel spectra after denoising and background subtraction. Spectrum 1 has a single peak at 726 cm$^{-1}$, matching the spectrum taken from the reference sample of isotopically pure $^{15}$NA. Spectrum 3 has a single band at 733 cm$^{-1}$, corresponding to the PESRS spectrum of $^{14}$NA. Spectrum 2 has a single peak at 730 cm$^{-1}$ that corresponds to the spectra of mixed molecules. Two distinct features could not be observed in the

PESRS spectra of these adenine isotopologues due to the spectral resolution of the hyperspectral SRS system (c.a. 14 cm$^{-1}$). Notably, however, spectra with single peaks at 726 and 733 cm$^{-1}$ were observed at multiple pixels (see **SI7**). These data allowed the statistical analysis described below.

A multivariate curve resolution (MCR) method was used for the statistical analysis of the PESRS spectra. The hyperspectral data (containing 40000 single-pixel spectra) was unmixed by MCR into concentration contributions of pure $^{14}$NA ($C_{14}$) and pure $^{15}$NA ($C_{15}$) spectra (**Fig. 4c**) and a relative concentration ratio of $^{14}$NA ($C_{14}/(C_{14}+C_{15})$), defined as the fraction of the average number of $^{14}$NA molecules contributing to the total signal, was thus determined. We selected 4172 of the total spectra acquired that displayed the desired Raman bands and had an intensity above a threshold value (maximum values > 0.01) were selected for this statistical analysis. The threshold requirement helped reduce inclusion of noise events and avoid the artificial counting of molecular events.[33] The histogram of relative contributions to the total signal produced by $^{14}$NA was obtained by counting the selected events. As shown in **Fig. 4d**, the histogram of intensity ratios has the dominant contribution from the single-molecule events at the ratio ≈ 0 and ≈ 1. Based on the Le Ru's statistical methodology,[33] for histograms with distributions like that shown in **Fig. 4d**, the edges of the histogram (events for the ratio ≈ 0 or ≈ 1) represent single-molecule events with a high probability. The obtained statistical results match the expected statistics prediction of "single-molecule PESRS model data" (**SI8**) well. Collectively, our results provide clear evidence that PESRS allows detection of single-molecule events for endogenous biomolecules having a cross-section as low as 10$^{-30}$ cm$^2$.

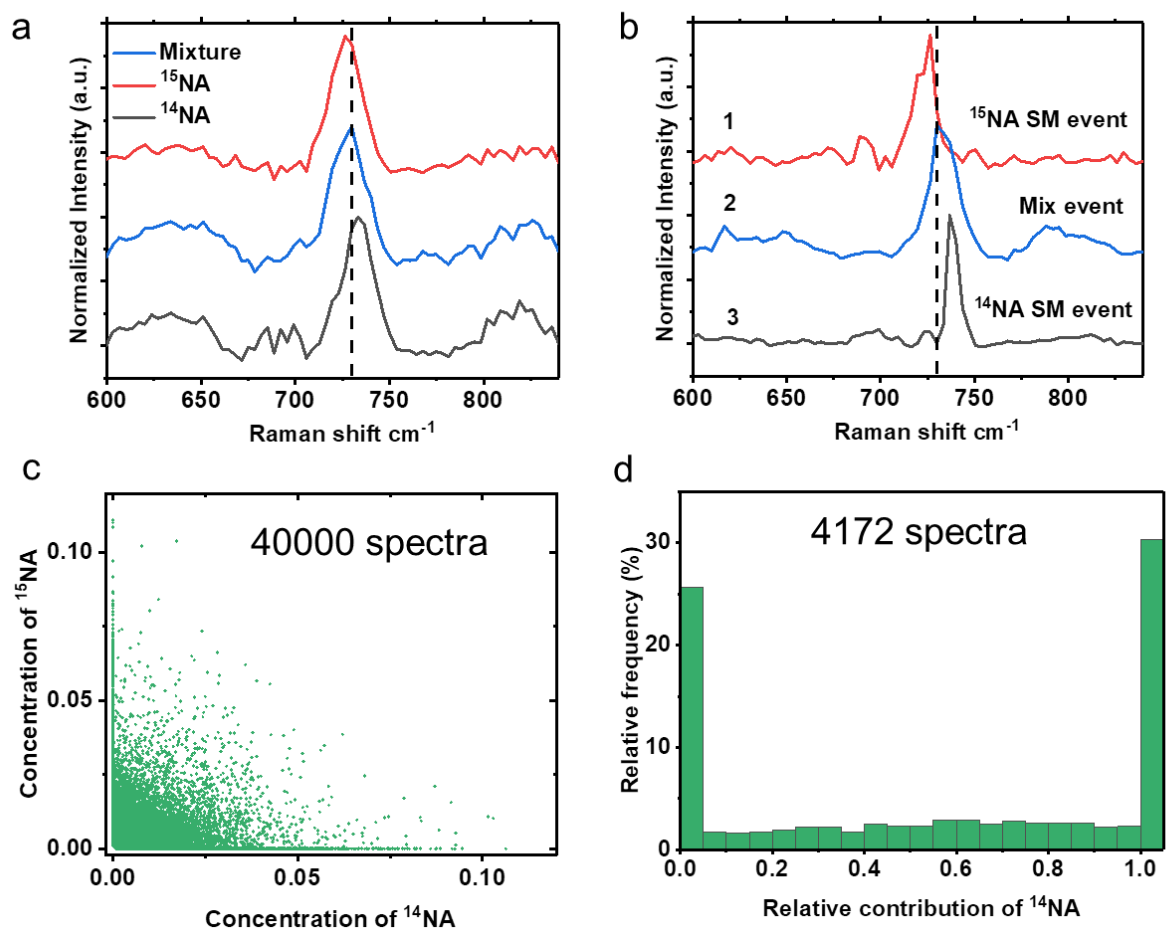

**Figure 4. Single-molecule sensitivity in PESRS.** (**a**) The ensemble-molecules PESRS measurements of pure $^{14}$NA, pure $^{15}$NA and their equimolar mixture. (**b**) Three representative single-pixel PESRS spectra showing a pure $^{15}$NA SM event (1), a mix event (2), and a pure $^{14}$NA SM event (3), which obtained from the 500 nM mixture sample. The vertical dash lines indicate the position of 730 cm$^{-1}$. (**c**) Concentration matrix coefficients obtained from MCR analysis of the hyperspectral imaging result sample (including 40000 single-pixel spectra) of the mixture. (**d**) Histogram of relative contribution of $^{14}$NA from the hyperspectral imaging result of the mixture sample. Selected 4172 single-pixel spectra with a desired Raman peak and above an intensity threshold were used. SM: single molecule.

**PESRS mapping of adenine generated from bacteria.** The investigation of dynamic living samples requires imaging at a high speed. Compared to SERS, the dramatically improved speed of PESRS microscopy makes it a potentially useful tool for imaging the chemical dynamics of a complex living system. To demonstrate such capacity, we studied the metabolic response of *S. aureus* to starvation, as shown in **Fig. 5a**. Following enrichment in a nutrient-rich environment, the *S. aureus* sample was washed and centrifuged in pure water. After 1 hour, 1 μL of a bacterial suspension was placed on the Au NPs-SiO$_2$ plasmonic substrate and once the water evaporated (~ 5 min) the PESRS signal was acquired. A control sample was similarly prepared but in contrast a PESRS spectrum was obtained without a 1-hour delay. PESRS spectra of *S. aureus* under starvation conditions for 1 hour are displayed in **Fig. 5b top**, and the observed spectra closely resemble the Raman spectrum of adenine. In contrast, the spectra of *S. aureus* obtained immediately (no waiting period) (**Fig. 5b bottom**) do not exhibit an adenine-like Raman band. These results are consistent with the SERS data (**SI9**).[37] These data imply that adenine, a purine degradation product, is secreted from *S. aureus* as a response to the no-nutrient, water-only environment.[37] As shown previously,[37] these molecular species are secreted from the bacterial cells under starvation conditions and appear most heavily concentrated in the pericellular region near the outer cell wall. **Fig. 5c** shows the PESRS image of starved *S. aureus* on the plasmonic substrate, which presents the distribution of the secreted adenine. The two representative single-pixel PESRS spectra of *S. aureus* are present in **Fig. 5d**. These results collectively demonstrate that PESRS has the potential for the study of the bacterial exogenous metabolome.

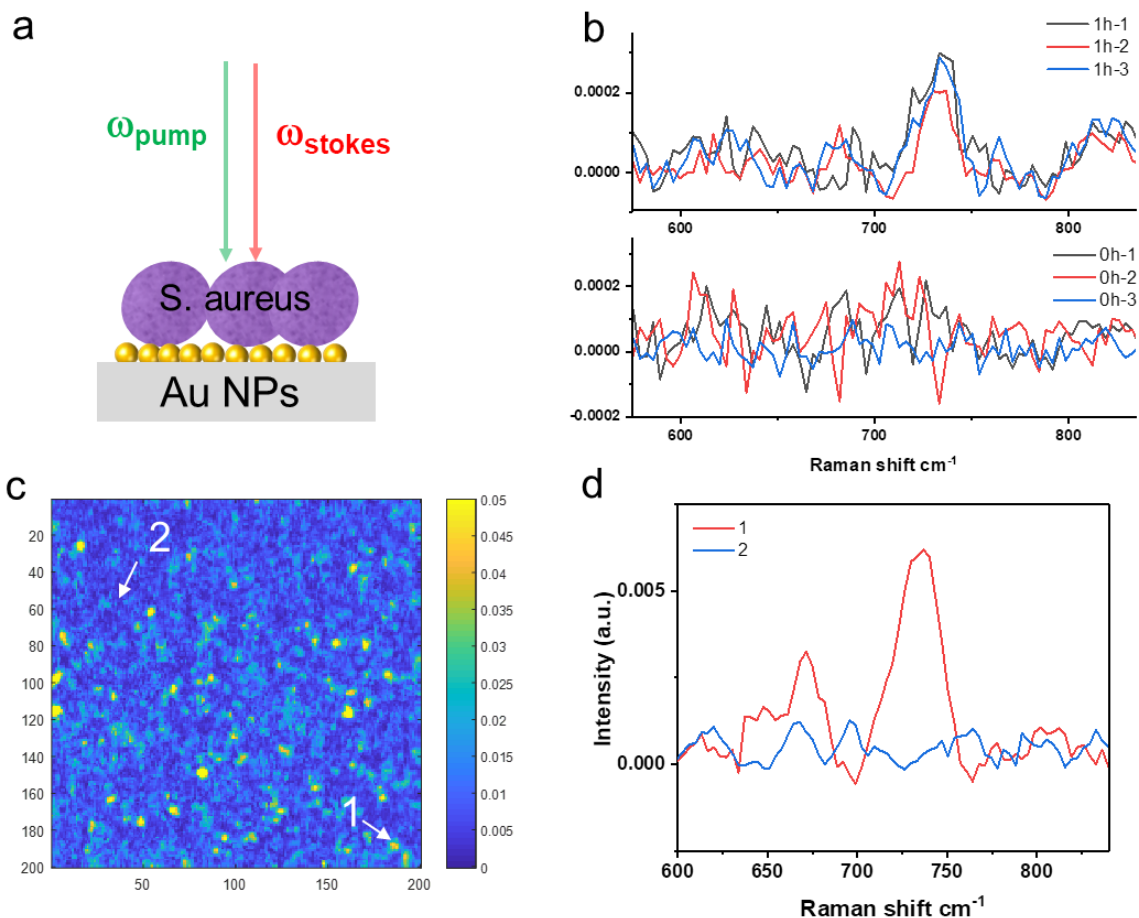

**Figure 5. PESRS mapping of adenine generated from stressed bacteria.** (**a**) Schematic. (**b**) Top: The PESRS spectra of *S. aureus* washed and kept in water for 1 hour. Bottom: the PESRS spectra of *S. aureus* obtained immediately after washing. Numbers 1, 2, 3 represent measurements at three locations on surface. No denoising applied. (**c**) Denoised PESRS image of starved *S. aureus* placed on the plasmonic substrate. The image area is 30 μm × 30 μm. (**d**) Single-pixel PESRS spectra of *S. aureus* on plasmonic substrate obtained from spot 1 and 2 indicated in panel c.

**Discussion**

Through plasmonic enhancement and hyperspectral recording, the detection sensitivity of SRS microscopy reached the single-molecule level. The Au plasmonic nanostructures provided an extraordinary SRS intensity enhancement relative to normal SRS of about $10^7$. Such large enhancements allow the detection of single molecules with a Raman cross-section as low as $10^{-30}$

cm$^2$. Single-molecule PESRS detection of adenine molecules was verified by a bianalyte method. A potential biomedical application of PESRS was demonstrated through mapping of adenine released from stressed bacterial cells.

In this work, single-molecule detection sensitivity in PESRS was achieved through a combination of several strategies. First, we created and employed a nanostructured Au substrate with a plasmon resonance peak overlapping with our pump and Stokes laser wavelengths. Second, we used chirped pJ laser pulses at 80 MHz repetition rate. Such pulse energy, which was 2 to 3 orders of magnitude lower than previous surface-enhanced femtosecond SRS work,[28, 29] significantly decreased potential photodamage.[29] The high repetition rate also allowed high-speed data acquisition. Third, a PLS method was used to distinguish the Raman peak from the broad non-Raman background in the hyperspectral dataset. Finally, a BM4D approach denoised a hyperspectral cube and allowed high-quality single-pixel spectra to be obtained for statistical analysis of single molecules events.

A portion of our recorded PESRS spectra show a dispersive vibrational line shape (**Fig. 1f**). Such line shapes were reported in previous surface-enhanced femtosecond SRS works.[28, 29, 42] Three possible explanations have been proposed: Fano-resonance of the molecule-nanostructure system,[42] interference between the PESRS signal and the plasmon enhanced aggregated NP emission,[43] and the effects of the complex character of the plamonic field amplitude.[44] The distribution and character of this PESRS line shape character will be described in subsequent work.

PESRS microscopy opens a new window for fast vibrational spectroscopic imaging of low-concentration molecules with high sensitivity. PESRS microscopy can sensitively detect metabolites secreted from a live cell. Thus, PESRS may be used to distinguish bacteria types and investigate metabolic changes linked to the development of microbial populations or to the exposure to antibiotics. In addition, this method can be extended to study the dynamic processes in surface chemistry, such as the mapping of the solid electrolyte interface membrane in a lithium cell or imaging the heterogeneity of catalyst.

Additional improvements may be envisioned for this technology. For example, the imaging speed can be further improved by using a multiplex SRS method,[15, 16] advanced delay tuning approach,[18] or a wide-field SRS system. Secondly, harnessing the rational-designed reproducible plasmonic nanostructure fabricated by lithographic methods, our method can pave the way for reproducible and quantitative molecular imaging platform. Such improvements will invoke the integration of coherent Raman imaging techniques and novel nanostructure designs to open new avenues towards ultra-sensitivity, ultra-fast, label-free chemical imaging.

## Methods

**Hyperspectral plasmon-enhanced stimulated Raman scattering microscope.** Figure S1 presents the scheme of a hyperspectral SRS microscope. Briefly, an 80 MHz tunable femtosecond laser (InSight DS+, Spectra-Physics) provided the pump (680-1300 nm) and Stokes (1040 nm) simulating fields. The Stokes beam was modulated by an acousto-optic modulator at 2.3 MHz. The pump and Stokes beams were spatially aligned and sent to an upright microscope with 2D galvo system for laser scanning. The spectral-focusing approach was used to obtained spectral domain information. In spectral focusing, the pump and Stokes pulses were equally stretched in time by glass rods to achieve a constant instantaneous frequency difference that drives a single Raman coherence. By delaying the pump pulses, a series of Raman shifts (80 channels) were generated. At a certain delay, all the laser energy was spectrally focused to excite a narrow Raman band. The laser powers (pump ~ 0.15 mW and Stokes ~ 0.15 mW at the sample) were sufficiently low so that good stability of the molecule and nanostructure was maintained during the experiments. A 60× water immersion objective (Olympus, NA=1.2) or a 40× water immersion objective (Olympus, NA=0.8) was used to focus pump and Stokes laser on a sample. An oil condenser (Nikon, NA=1.4) was used to collect the laser light in the forward direction. A 1000 nm short-pass filter (Thorlabs) blocked the Stokes laser before a photodiode with a lab-built resonant amplifier. A lock-in amplifier demodulated the stimulated Raman loss signal from the pump beam detected by the photodiode. An XY scanner with a 150 nm step scanned the sample, and a PESRS hyperspectral data cube (200 × 200 pixels, 80 Raman channels) was recorded with a 10 μs dwell time per pixel. The PESRS imaging speed was c.a. 1 min per data cube. The image size (30 μm × 30 μm) and pixel dwell time were the same in all experiments.

An epi-detected SRS microscope was built for PESRS detection on non-transparent plasmon-enhanced substrates. Before the microscope, a quarter wave plate was placed after a polarizing beam splitter to change the polarization of excitation and back-reflected laser light by 90°. In this way, the polarizing beam splitter allowed forward light to pass through and the stimulated Raman loss signals were reflected into a photodiode to achieve epi-PESRS imaging (**Fig. 3a**).

**Background reduction in PESRS.** A raw PESRS spectrum contains a large background signal from the photothermal effect, cross-phase modulation, and transient absorption. Cross-phase modulation originates from the optical Kerr effect, and the transient absorption and photothermal effect are due to the plasmonic resonance of the Au NPs. We minimized the background arising from non-Raman processes by using a larger numerical aperture (NA=1.4) lens for signal collection to reduced cross-phase modulation and photothermal effect. Moreover, a MHz-frequency modulation was used to further diminish the photothermal effect. With those approaches, we successfully observed a PESRS signal in the presence of a strong background.

**Background subtraction in a PESRS spectrum**. An adaptive iteratively reweighted penalized least squares (airPLS) algorithm, developed by Zhang et al.[45], was employed to subtract the baseline from the raw PESRS spectrum.

**Denoising of a PESRS hyperspectral data**. Firstly, we used a BM4D denoising algorithm, developed by Maggioni and Foi,[39, 40] to process the raw PESRS hyperspectral data cube. The BM4D algorithm relies on the so-called grouping and collaborative filtering paradigm. A 3D imaging block (x-y-λ) was stacked into a 4D data array, which was then filtered. Thus, BM4D leveraged the spatial and spectral correlation of a hyperspectral data cube both at the nonlocal and local level. Then, we used the airPLS algorithm to subtract the background of a hyperspectral data cube pixel by pixel. In addition, the image was plotted by the peak area of the desired Raman band.

**Statistical analysis of single-molecule events.** Before the statistical analysis of single-molecule events, the PESRS hyperspectral data was denoised and corrected baseline as described. A multivariate curve resolution (MCR) algorithm, developed by Tauler and Juan et al.,[46, 47] was used to extract the concentration maps of the two isotopically related molecules in the mixed hyperspectral dataset. Because $^{14}$NA and $^{15}$NA have the same Raman cross-section, we used the normalized spectra separately obtained from pure $^{14}$NA and $^{15}$NA samples as the initial estimation of the pure spectra. The constraints implemented during the optimization step were non-negativity for the concentration and spectrum. The outputs of the MCR treatment were the pure concentration maps ($C_{14}$, $C_{15}$) of $^{14}$NA and $^{15}$NA, respectively. Then, the 4172 spectra whose maximum values appeared at the desired wavenumber range were selected and an intensity threshold (maximum values > 0.01) was defined for removal of noise events. The histogram of the relative contribution of $^{14}$NA ($C_{14}/(C_{14}+C_{15})$) were plotted and the edges of the histogram, $C_{14}/(C_{14}+C_{15}) \approx 0$ or $\approx 1$, were considered as the single molecules events. All data were processed by MATLAB.

**Substrate preparation.** The Au NPs colloid was prepared according to the classical citrate reduction method,[48] and resulted in particles with a diameter of ~ 50 to 60 nm, as shown in **SI10**. The 0.5 mL of 0.01% (g/ml) colloidal suspension was concentrated to 2-4 µL by centrifuging, which was then added to the 10 µL adenine solution. The adenine solution induced the aggregation of Au NPs. The aggregated Au NPs were dropped on a cover glass, followed by vacuum drying to obtain a substrate for PESRS detection.

Au NPs-SiO$_2$ plasmonic substrates were prepared as described previously.[41] Immobilized clustered aggregates of 80 nm Au NPs were grown on a SiO$_2$ chip. A 10 µL adenine solution was dropped on the Au NPs-SiO$_2$ plasmonic substrates and samples was ready for PESRS measurement after adenine solution dried under air ambient (~ 5 mins). High-purity water (Milli-Q, 18.2 MΩ·cm) was used throughout the study.

**Bacteria sample preparation**: Bacteria were harvested during the log phase. Culture growth media was removed from the bacterial samples by centrifugation, and washing four times with 2 mL of deionized Millipore water. The bacterial pellet was suspended in 0.25 mL of water, and 1 µL of the resulting bacterial suspension was dropped and dried onto the Au NPs-SiO$_2$ plasmonic substrate. Samples were dried onto the Au substrate either immediately or after 1 hour in order to demonstrate the effect of the starvation stress response.

Additional experimental results and data analysis are available in the supplementary information.

**Data availability.** The data that support the plots within this paper and other findings of this study are available from the corresponding author upon reasonable request.

**Acknowledgements**: This work was supported by a Keck Foundation grant to J.X.C., Xiamen University postdoctoral fellowship to C. Zong supported by NSFC (2163000117, 21621091 and 21790354), and MOST (2016YFA0200601) grants to B.R. L.D.Z acknowledges the support of NSF (CHE-1609952). We thank Jiayingzi Wu for assistance with UV-Vis detection, and Kai-Chih Huang for his valuable suggestion that greatly improved the experimental performance.

**Author contributions:** Experiments were designed by J.X.C, C. Zong, and L.D.Z.. The PESRS experiments were conducted by C. Zong. Data analysis was executed by C. Zong with the help of H.N.L. P.R. prepared the Au NPs-SiO$_2$ substrate and the bacteria sample. Y.M.H performed the SEM study and UV-Vis detection. H.N.L, C. Zhang, and C. Zong developed the hyperspectral SRS instruments. C. Zong wrote the manuscript, revised by J.X.C., C.Y., B.R., and L.D.Z.. All authors commented on the manuscript.

**Competing interests**

The authors declare no competing financial interests.

# Supporting information

# Plasmon-enhanced Stimulated Raman Scattering Microscopy with Single-molecules Detection Sensitivity

Cheng Zong, Ranjith Premasiri, Haonan Lin, Yimin Huang, Chi Zhang, Chen Yang, Bin Ren, Lawrence D. Ziegler, and Ji-Xin Cheng

**SI1. Scheme of a hyperspectral plasmon-enhanced stimulated Raman scattering microscope.**

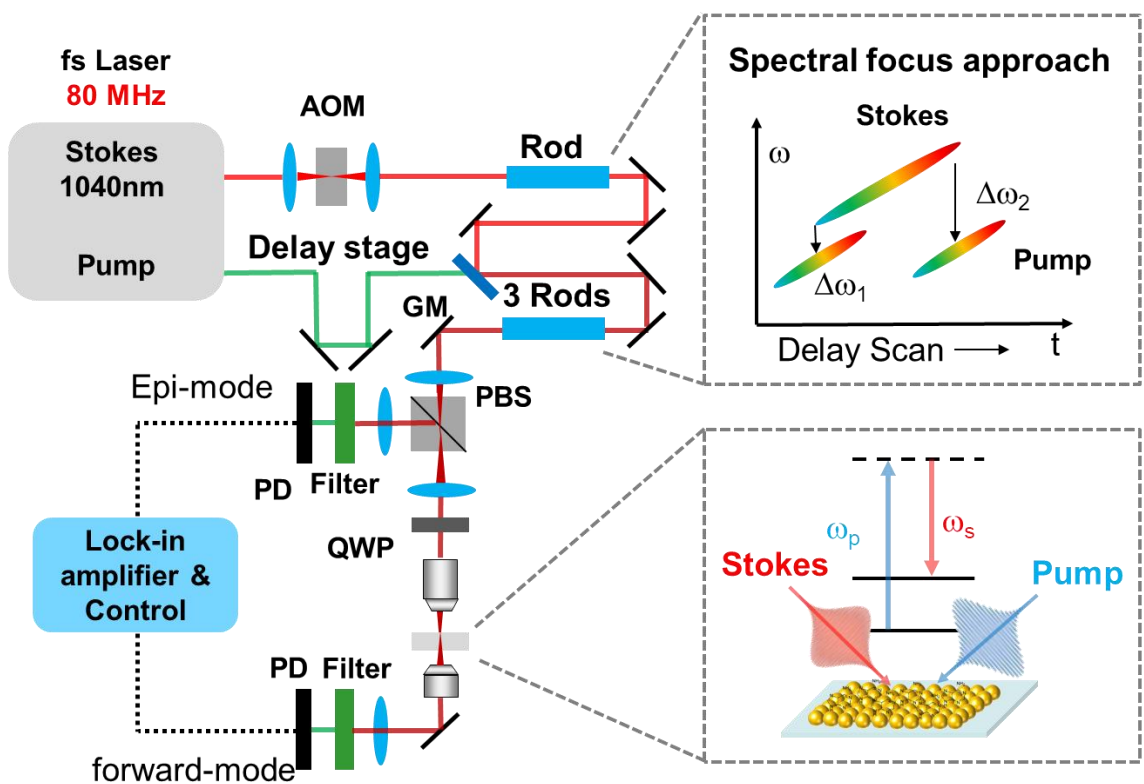

**Figure S1.** The scheme of a hyperspectral plasmon-enhanced stimulated Raman scattering microscope. Laser system: 80 MHz tunable femtosecond laser. AOM: acousto-optic modulator. GM: 2D galvo mirror. PBS: polarizing beam splitter; QWP: quarter wave plate; PD: photodiode. For spectral focusing, three rods were used in combined path and one rod was used in Stokes path. In this way, the pump and Stokes pulses were chirped to achieve a constant instantaneous frequency difference that drives a single Raman coherence. A series of Raman shifts were generated by scanning the delay stage.

**SI2. The SERS spectrum of adenine adsorbed on Au NPs aggregation substrate and the Raman spectrum of adenine solution.**

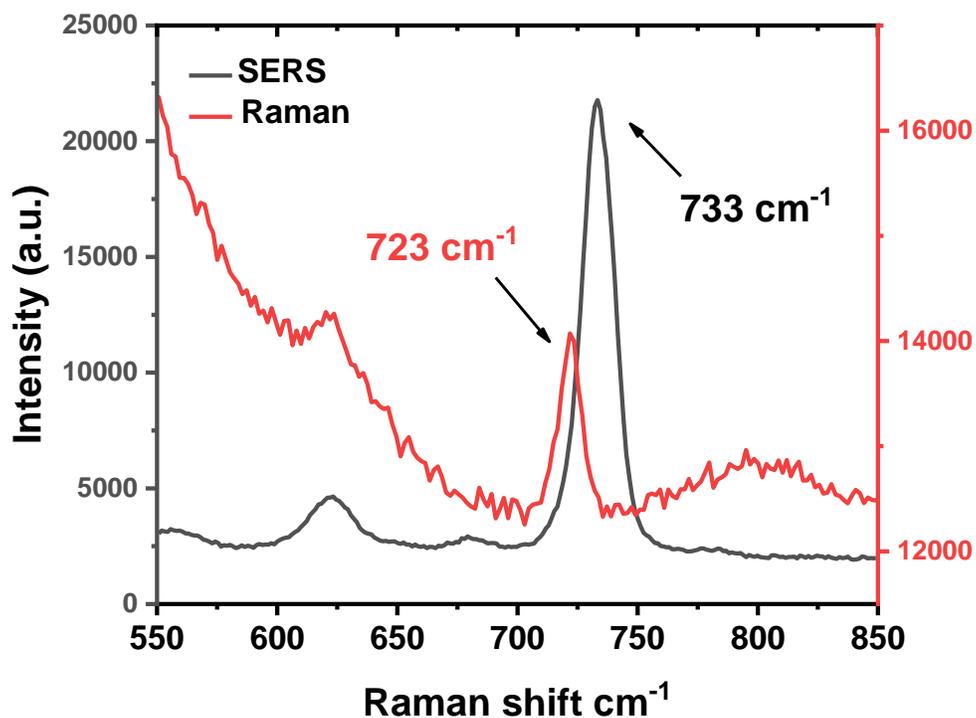

**Figure S2.** The SERS spectrum (black) of adenine adsorbed on Au NPs aggregation substrate (5 mM in solution) has a peak at 733 cm$^{-1}$. The Raman spectrum (red) of 5 mM adenine solution has a peak at 723 cm$^{-1}$. The SERS spectrum was recorded with 5 s integration time, with a 50× objective and a 0.5 mW laser power at 785 nm. The Raman spectrum was recorded with 30 s integration time, with a 40× objective and an 80 mW laser power at 532 nm. This 10 cm$^{-1}$ blue shift is due to the formation of metal-adenine complex.[1].

**SI3. Dependence of plasmon-enhanced stimulated Raman scattering (PESRS) signal on pump and Stokes laser power.**

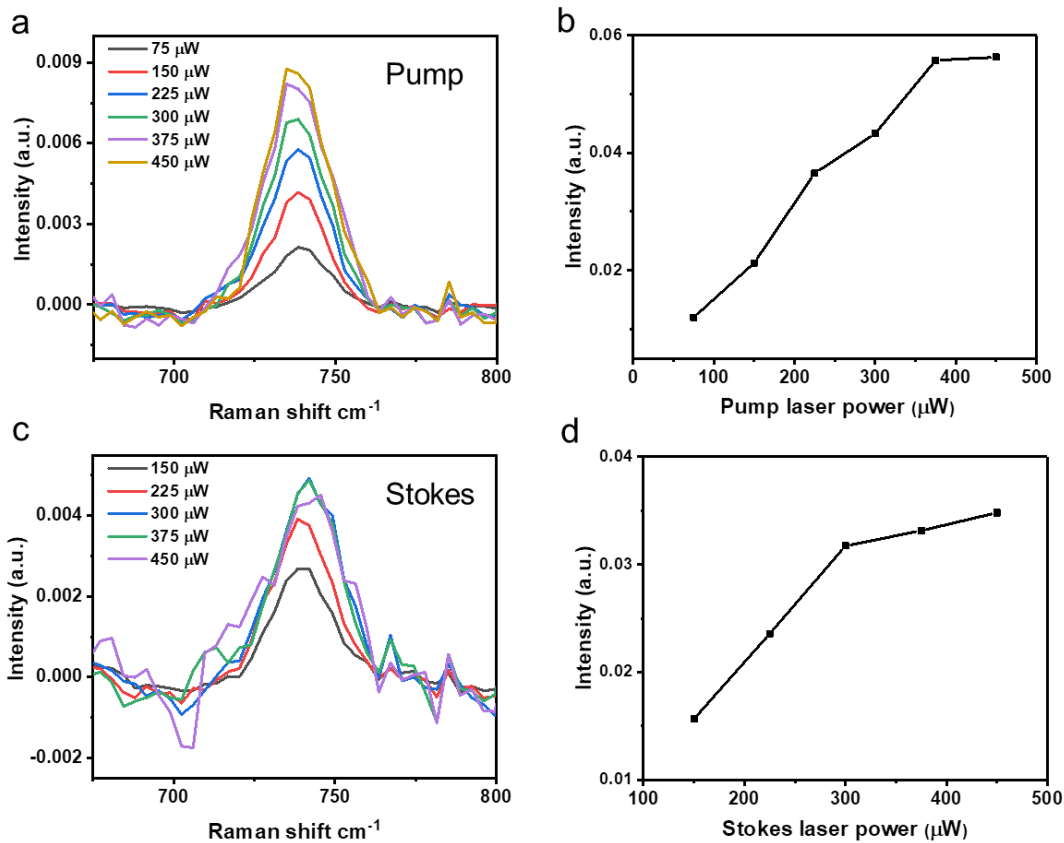

**Figure S3**. Power dependence of PESRS signal. (a) The pump power dependent PESRS spectra of adenine adsorbed on Au NPs aggregation substrate at the same position. The Stokes power was kept at 150 μW. (b) The peak area at 733 cm$^{-1}$ of adenine obtained from (a) vs. pump power. (c) The Stokes power dependent PESRS spectra of adenine adsorbed on Au NPs aggregation substrate at the same position. The pump power was kept at 150 μW. (d) The peak area at 733 cm$^{-1}$ of adenine obtained from (c) vs. Stokes power. The power value was the power at the sample. The PESRS signal has a linear relationship with the pump and Stokes power under a saturation power threshold (the sum of two laser power is c.a. 600 to 650 μW). The spectral features are similar. All laser powers, used in whole PESRS experiments, were kept below this threshold to avoid photodamage of samples.

**SI4. SERS and PESRS spectra of Rhodamine 800 (RH800) and 4-mercaptopyridine (Mpy).**

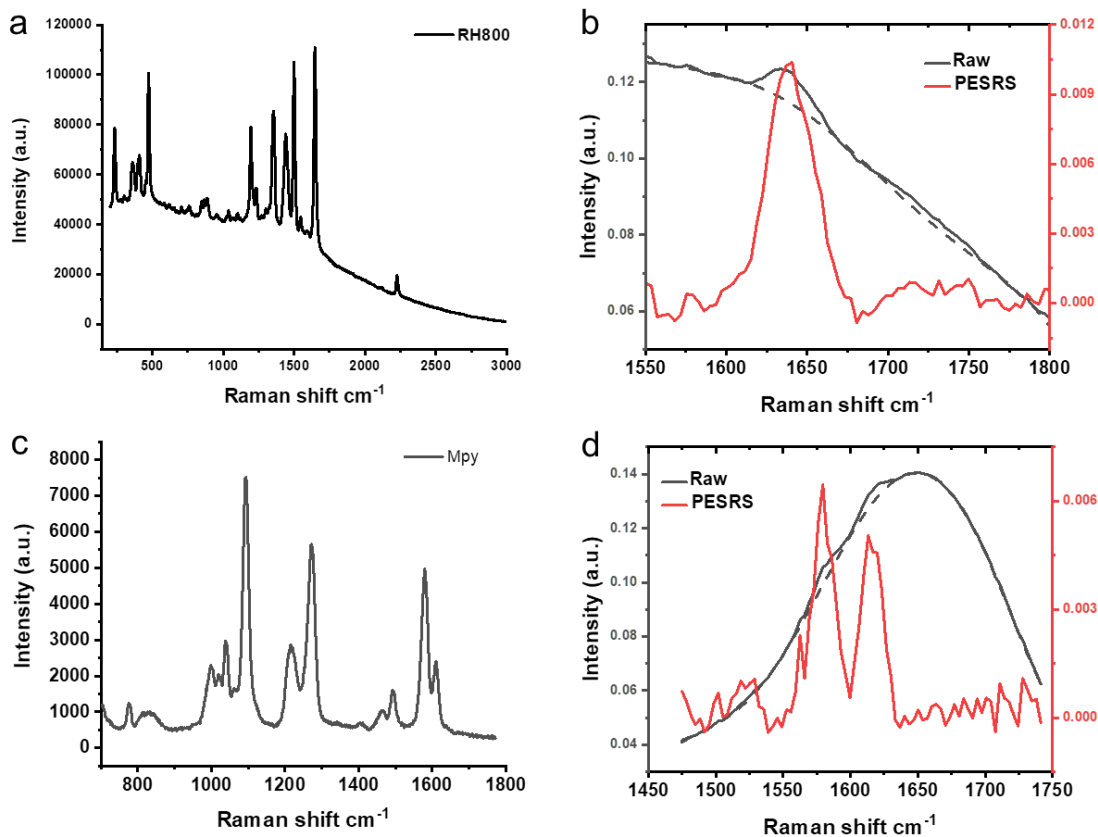

**Figure S4.** (a) The SERS spectrum of RH800 adsorbed on Au NPs. The spectrum were recorded with 10 s integration time with a 50× objective and a 3 µW laser power at 785 nm. (b) Original PESRS spectrum (black solid) and fitted background (black dash) of RH800 adsorbed on Au NPs. The background-subtracted PESRS spectrum (red) of RH800. 75 µW pump laser (888 nm) and 50 µW Stokes laser were used. 10 µL of 85 µM RH800 was dropped into centrifuged Au NPs sol and was dry in vacuum. (c) The SERS spectrum of Mpy adsorbed on Au NPs. The spectrum were recorded with 2 s integration time with a 50× objective and a 300 µW laser power at 785 nm. (d) Original PESRS spectrum (black solid) and fitted background (black dash) of Mpy adsorbed on Au NPs. The background-subtracted PESRS spectrum (red) of Mpy. 150 µW pump laser (891 nm) and 150 µW Stokes laser were used. 10 µL of 5.7 mM Mpy was dropped into centrifuged Au NPs sol and was dry in vacuum. This result indicated that our method can obtain PESRS spectra from a variety of molecules.

**SI5. Representative single-pixel PESRS spectra of adenine from a single imaging indicate good reproducibility.**

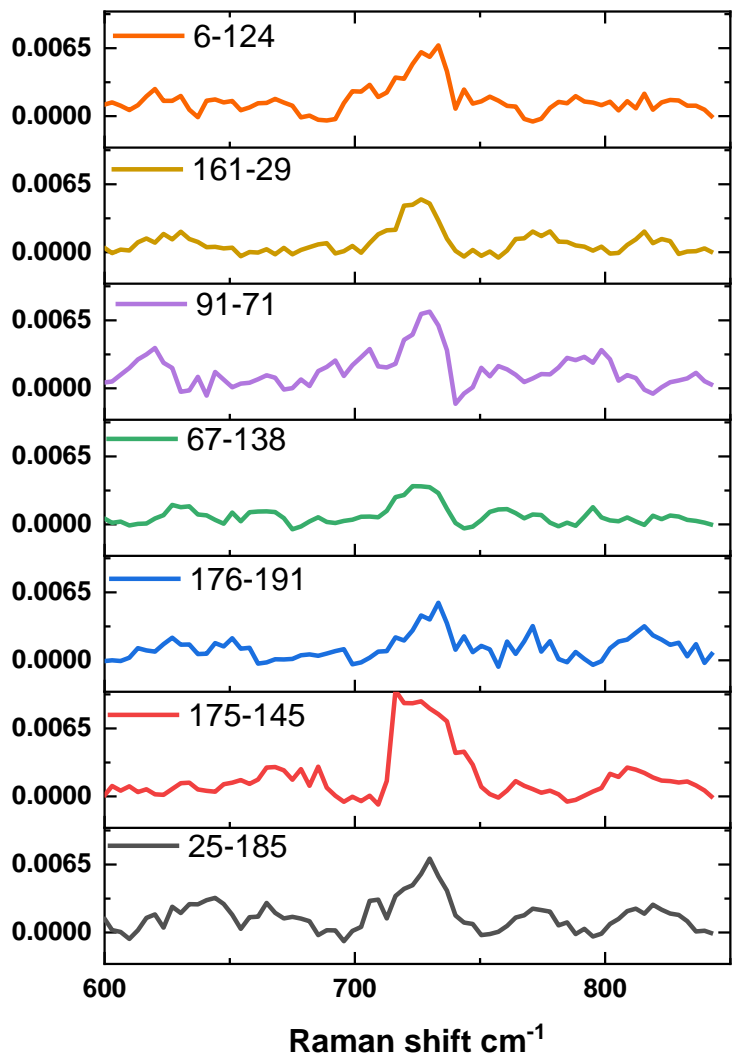

**Figure S5.** Representative single-pixel PESRS spectra of adenine, obtained from aggregated Au NPs substrate (the same hyperspectral data cube in Figure 2). The labels of each spectrum indicate the X-Y pixel values.

**SI6. The estimation of local enhancement factor of PESRS.**

In SRS, the equation of the intensity of SRS signal is shown as following[2]:

$$I_{SRS} \propto N * \sigma * P * S$$

Where, $I$ is the intensity, $N$ is the number of molecules under the laser spot, $\sigma$ is the molecular Raman scattering cross-section. $P$ is the pump laser power. $S$ is the Stokes laser power. To estimate the enhancement factor (EF) of PESRS, we use the following equation to calculate the power- and concentration- averaged EF between PESRS and SRS.

$$EF = \frac{I_{PESRS}}{I_{SRS}} * \frac{N_{SRS}}{N_{PESRS}} * \frac{P_{SRS}}{P_{PESRS}} * \frac{S_{SRS}}{S_{PESRS}}$$

As shown in Figure S6, the SRS spectrum (average 200 × 200 pixel area spectra) of 5 mM adenine solution was measured at the power of 15 mW (Pump) and 100 mW (Stokes) and the PESRS spectrum (average 3×3 pixel area spectra) of adsorbed adenine was measured at the power of 0.5 mW (Pump) and 0.5 mW (Stokes). We assumed the size of the laser spot was 500 nm, the size of adenine was 0.5 nm² per molecule. Au NPs was a monolayer under the laser spot and a monolayer adenine adsorbed on Au NPs surface. The number of molecules in detection volume (c.a. 30 μm × 30 μm × 1 μm = 900 μm³) was about $2.7 \times 10^9$ for SRS detection. The number of molecules on the surface can be estimated as about $1.6 \times 10^6$ for PESRS. In this way, the local enhancement factor of PESRS was about $7 \times 10^7$.

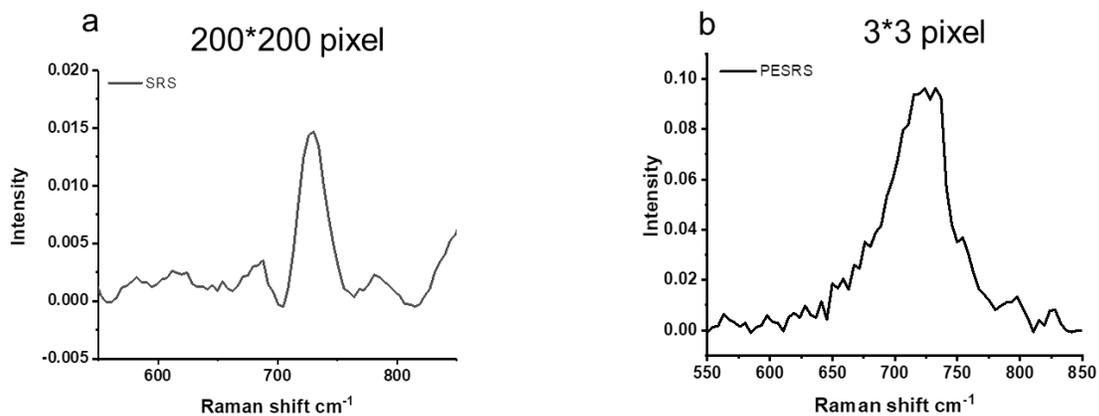

**Figure S6**. (a) The SRS spectrum of 5 mM adenine solution (average 200×200 pixel area spectra). Pump power: 15mW, Stokes power: 100 mW. (b) The PESRS spectrum of adsorbed adenine on Au NPs-SiO$_2$ substrate (average 3×3 pixel area spectra). Pump power: 0.5 mW, Stokes power: 0.5 mW.

**SI7. The representative single-molecule events of adenine in a PESRS image.**

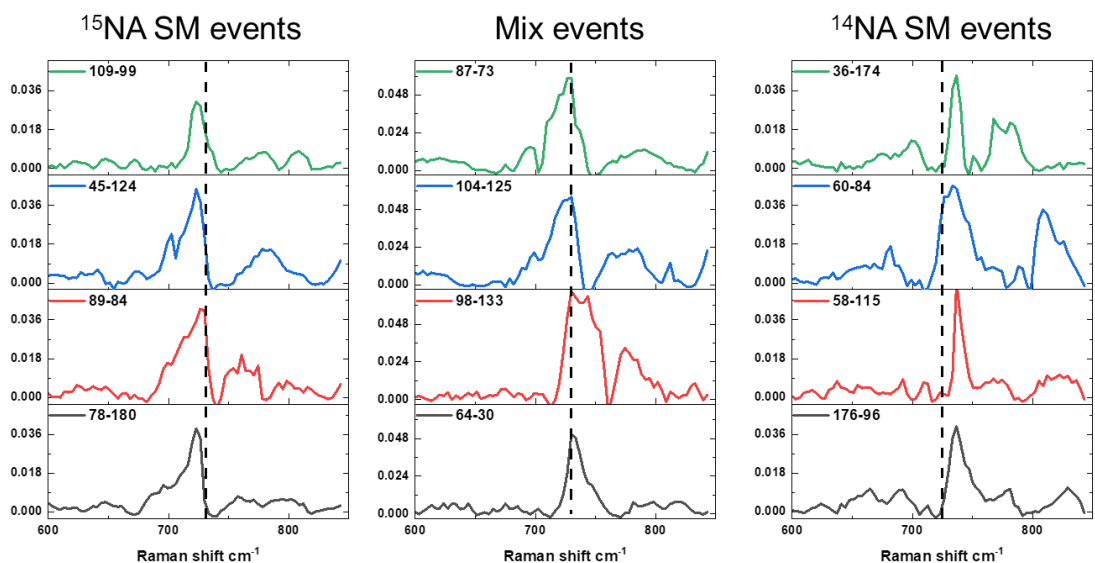

**Figure S7.** The representative single-molecule (SM) events of adenine. Left column: pure $^{15}$NA single-molecule events, Right column: pure $^{14}$NA single-molecule events. Middle: mix events. The vertical dash lines indicate the position of 730 cm$^{-1}$. Fluctuations in intensities and spectral feature were observed, most likely due to the variation of molecular orientation in hot spots. The fluctuations are a signature of single-molecule events.

**SI8. The simulated single-molecule PESRS data**

Here, we introduced a model example to describe the statistics of single-molecule PESRS signal in a hot spot. This model was based on the previous single-molecule SERS mode developed by Eric Le Ru, PG Etchegoin, et at.[3] First, we used the boundary element method approach[4-6] to calculate a local electric field distribution on a representative hot spot. Fig S8a presented the simulated local electric field distribution of the representative hot spot (a dimer formed by two 60 Au NPs with a 1 nm gap). Then, we assumed that we had a certain number of molecules of two isotopic adenines ($N_{14}$ and $N_{15}$) on the hot spot. We generated random locations of molecules in the hot spot and every molecule felt a corresponding local electric field in the hot spot.[7] Then, we calculated the total intensity produced by each type of molecules ($I_{14}$ and $I_{15}$) by summing over the corresponding intensity of every molecules. Because the Raman cross section of $^{14}NA$ and $^{15}NA$ were the same, the ratio of $I_{14}/(I_{14}+I_{15})$ was also the ratio of the average number of $^{14}NA$ contributing to the signal. We repeated this process for many times (as a large number of events) and obtained the histogram for relative contribution of $^{14}NA$. Figure S8b-d shows the the simulated results for repeating 9000 events with $N_{14}=N_{15}=10$, 100, and 1000 molecules, respectively. Fig S8b shows the single-molecule regime, $N_{14}=N_{15}=10$. There was a small number of mixed-signal events. Events were dominant by the ratio ≈ 0 and ≈ 1. This distribution indicated the event (ratio≈ 0 or 1) can be contributed to single-molecule events with a high probability. The many-molecules regime, $N_{14}=N_{15}=1000$. The histogram (Fig S8d) looks like a Gaussian distribution center at the ratio=0.5. With $N_{14}=N_{15}=100$, Fig S8c represents a transitional regime between the previous two. Compared with Fig S8a, more mix events could be observed in Fig S8c. As shown in Fig S8e, our 500 nM experimental result matched well the simulated result of 10 molecules, and the relative frequency of single-molecule events is larger than the simulation result of 10 molecules. This result indicated that our experiment result has achieved a single-molecule detection regime. Moreover, in Fig S8f, we compared the histograms of the relative contribution of $^{14}NA$ in the 500 nM and 5 μM of mixture sample. The molecule events of 5 μM mixture experiment results are also dominated by the single-molecule events. While the relative frequency of mix events in 5 μM sample is larger than that in 500 nM sample, which is matched well with our expectation and the simulation result that more mix events could be observed in the higher concentration sample.

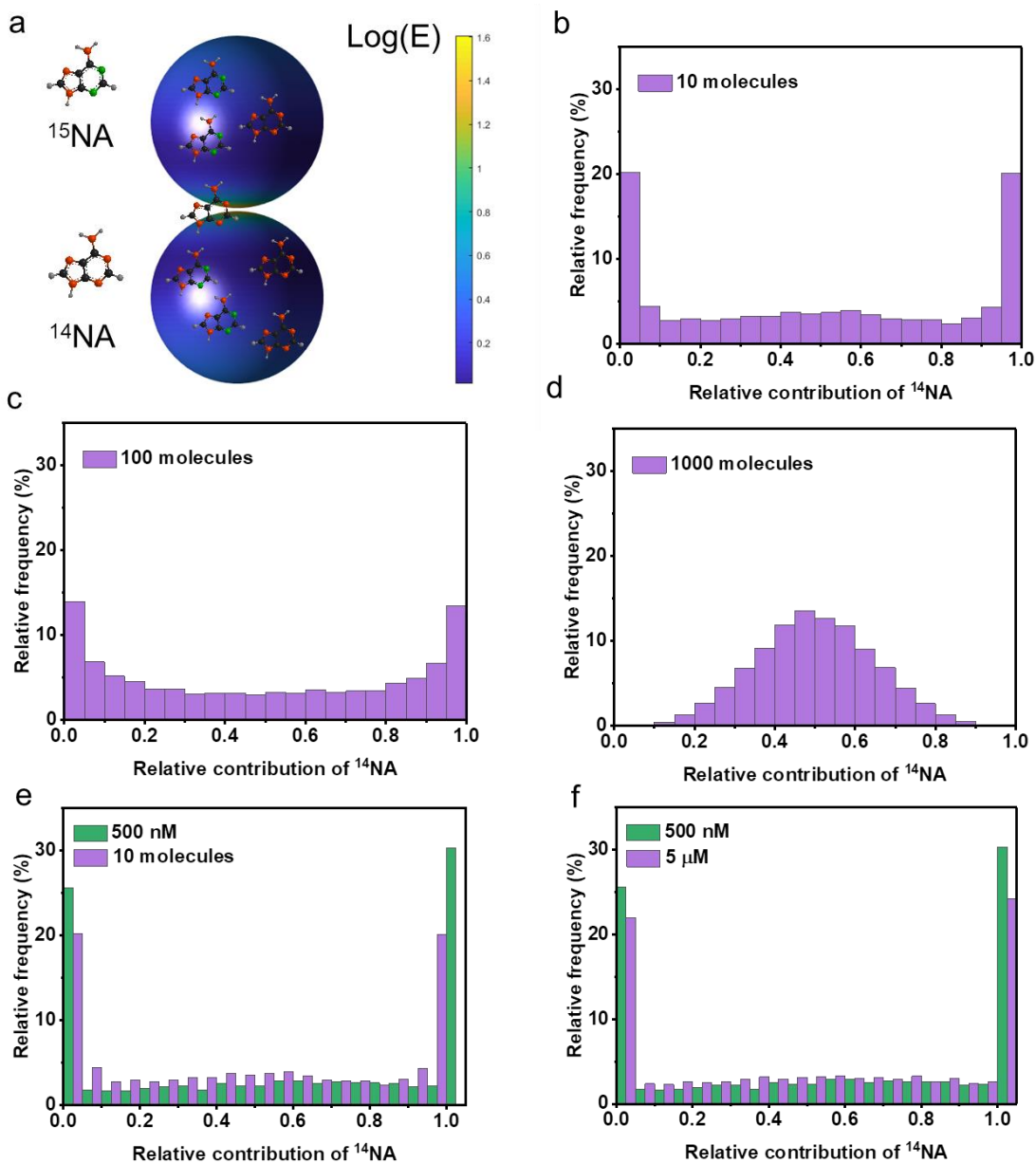

**Figure S8** (a) The local electric field distribution of the hot spot. The hot spot was formed from two 60 Au NPs with a 1 nm gap. Each molecule adsorbed on a certain location can experience a corresponding local electric field. (b-d) The histograms of the relative contribution of $^{14}$NA in the simulated results repeating 9000 events with $N_{14}=N_{15}=10$ (b), 100 (c), 1000 (d) molecules, respectively. (e) The histograms of the relative contribution of $^{14}$NA in the 500 nM mixture sample and in simulated results of $N_{14}=N_{15}=10$; (f) The histograms of the relative contribution of $^{14}$NA in the 500 nM and 5 μM mixture samples.

**SI9. SERS spectra of starved *S. aureus* and non-starved *S. aureus*.**

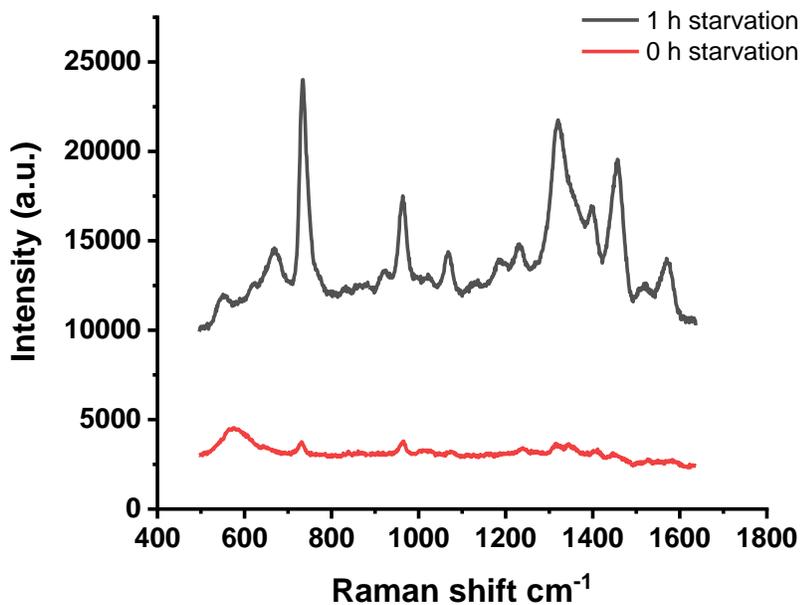

**Figure S9.** SERS spectra of starved *S. aureus* (black) and non-starved *S. aureus* (red). After 1 hour starvation in pure water, the SERS spectrum of *S. aureus* closely resembled the SERS of adenine. This result indicates that the adenine appeared at the outer layer of *S. aureus* and in the extracellular metabolome as resulting from the bacterial cell stress response to the no-nutrient, water-only environment. The spectra were recorded with 1.0 s integration time with a 50× objective and a 1.0 mW laser power at 785 nm.

**SI10. SEM images of Au NPs colloid.**

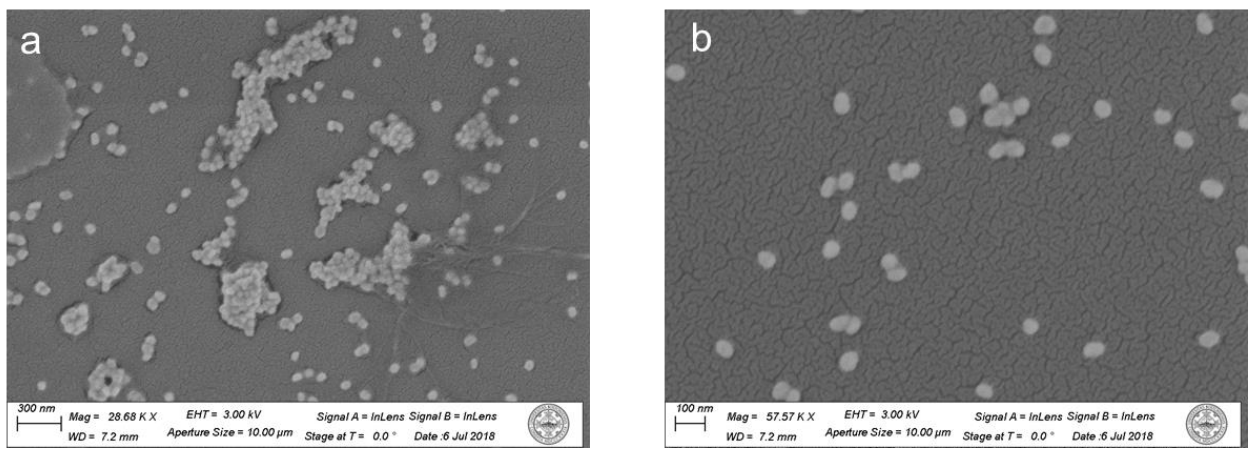

**Figure S10.** SEM images of Au NPs colloid. The scale bar in (a) is 300 nm, and in (b) is 100 nm. The size of Au NPs is around 50 to 60 nm.